# Forward and Inverse Modelling Approaches for Prediction of Light Stimulus from Electrophysiological Response in Plants

Shre Kumar Chatterjee[a], Sanmitra Ghosh[a], Saptarshi Das[a,*], Veronica Manzella[b,c], Andrea Vitaletti[b,c], Elisa Masi[d], Luisa Santopolo[d], Stefano Mancuso[d], and Koushik Maharatna[a]

a) School of Electronics and Computer Science, University of Southampton, Southampton SO17 1BJ, United Kingdom.
b) WLAB S.r.l., via Adolfo Ravà 124, 00142, Rome, Italy.
c) DIAG, SAPIENZA Università di Roma, via Ariosto 25, 00185, Rome, Italy.
d) Department of Horticulture, University of Florence, viale delle Idee 30, 50019, Sesto Fiorentino, FI, Italy.

**Authors' Emails:**

skc105@ecs.soton.ac.uk (S.K. Chatterjee), sg5g10@ecs.soton.ac.uk (S. Ghosh), sd2a11@ecs.soton.ac.uk (S. Das*), veronica.manzella@gmail.com (V. Manzella), andrea.vitaletti@w-lab.it (A. Vitaletti), elisa.masi@unifi.it (E. Masi), luisa.santopolo@unifi.it (L. Santopolo), stefano.mancuso@unifi.it (S. Mancuso), km3@ecs.soton.ac.uk (K. Maharatna)

**Phone no:** +44(0)7448572598, **Fax:** 02380 593045

**Abstract:**

In this paper, system identification approach has been adopted to develop a novel dynamical model for describing the relationship between light as an environmental stimulus and the electrical response as the measured output for a bay leaf (*Laurus nobilis*) plant. More specifically, the target is to predict the characteristics of the input light stimulus (in terms of on-off timing, duration and intensity) from the measured electrical response – leading to an inverse problem. We explored two major classes of system estimators to develop dynamical models – linear and nonlinear – and their several variants for establishing a forward and also an inverse relationship between the light stimulus and plant electrical response. The best class of models are given by the Nonlinear Hammerstein-Wiener (NLHW) estimator showing good data fitting results over other linear and nonlinear estimators in a statistical sense. Consequently, a few set of models using different functional variants of NLHW has been developed and their accuracy in detecting the on-off timing and intensity of the input light stimulus are compared for 19 independent plant datasets (including 2 additional species *viz. Zamioculcas zamiifolia* and *Cucumis sativus*) under similar experimental scenario.

1. Introduction

It was discovered by Burdon-Sanderson in 1873, that plants exhibit bioelectrical activity [1]. In 1926, J.C. Bose isolated the vascular bundles of a fern to show that physiological events, such as those present in animal nerves, triggered excitation which travels as electrical signals within plants [2]. In the recent years, it has been observed by botanists that electrical signals, naturally occurring in plants are generated due to different physiological processes induced by external stimuli [3]. Such stimuli could be in the form of change of light intensity [4][5][6], temperature [7][8], humidity [9], introduction of gas [10], mechanical wounding [11] etc. A detailed analysis of plant electrophysiological systems is documented in [12]. Although there were a few attempts to understand the nature of such electrical signals using signal processing techniques [13–20], rigorous models correlating the characteristics of such signals with the externally applied stimuli are yet to be developed.

When compared to the wealth of existing knowledge of animal electrophysiology, in lieu of stimulus induced response phenomenon, the same for plants are still evolving. Consequently there exists little knowledge about how the electrical response in plants changes in a stimulus-specific way and how the changes of electrical responses could be quantitatively associated with the stimulus. A model describing such behaviour will allow us to quantitatively associate different environmental parameters to the electrophysiological responses of different plants. Thus such a model will provide invaluable information which will augment the existing technologies present in agriculture and environment monitoring applications. From a system theoretic perspective such a model could be formulated by considering plants as dynamical systems and observing their input-output (stimulus-electrical response) relationships. The mechanisms of light-stimulus induced electrical signal generation in plants are quite well-researched in the plant physiology community. Roelfsema *et al.* [21] reported the findings of three physiological states of the guard cells, which are electrically isolated from other plant cells. These states were found to be far-depolarized, depolarized and hyperpolarized. The depolarized guard cells were found to extrude potassium ($K^+$) ions through the outward rectifying channels whereas the hyperpolarized cells let in $K^+$ ions through inward rectifying channels. The guard cells were reportedly switching from depolarized to hyperpolarized state upon incidence of light and vice versa for a light to darkness transition. The average membrane potential was found to be -41mV during depolarization and -112 mV during hyperpolarization as per the study by [21]. Such a mechanistic model essentially describes the electrical potential generation process in single plant cell whereas the collective behaviour of such electrical potentials could be observed in a realistic experimental scenario. In order to model the global effect of several such electrical signals generated by a tissue (several such cells) in a plant due to external light excitation, we have tested several linear and nonlinear statistical estimators to best describe the plant electro-physical using a dynamic model.

In principle, modelling input-output relations of any dynamical system could be done in two ways. The first method involves a mechanistic approach where in detailed understanding of the characteristics of physical interactions between the system's components with the input stimuli are exploited. The second method involves, considering the system as a black-box and statistically formulating the functional relationship by observing the output response to a given set of input stimuli. While the first approach is mostly desirable as it gives a complete understanding of the internal operation of the system, the second approach, which is often known as system identification, is more suitable for developing a working solution when the knowledge about the interactions between the system components is not adequate. For modelling the electrical response of plants under external stimuli, from the perspective of the level of existing knowledge, the system identification approach appears to be most appropriate since the knowledge about internal physiological processes in plants given a set of input stimuli is still not adequate.

In order to explain the plant electrical response due to light stimulus from a fundamental perspective several mechanistic models have been proposed by plant scientist. Models proposed in [22–25] and [26] explain the generation of an action potential and variation potential respectively. In [3], several models describing the underlying generation process of action potentials and variation potentials in plants are reviewed. These models essentially describe the generation of different ion currents across a cell and how these currents lead towards a transient depolarization of the membrane potential. Furthermore, the interaction of these currents with the transmembrane voltage is described using either a linear or Ohmic or a nonlinear relationship such as the Goldman-Hodgkin-Katz equation [27]. The stimulus is related in these models through stimulus induced calcium current that triggers further ionic currents. Also it is worth noting that none of these models quantify the time course of the stimulus intensity. It is for the purpose of modelling the phenomenon of stimulus driven electrical response. Thus an attempt to construct a mechanistic model is futile, as the relationship between the electrical responses recorded from the surfaces of plants and that from the cell is not well understood. Furthermore, morphologically a surface recording of the stimulus driven electrical response may show traces of action potential or a variation potential or a combination of both [28]. Thus, a mechanistic model which might describe and characterise all these traits would induce complexities that might hinder the applicability of the model. Motivated by these constraints imposed on mechanistic modelling, we have taken a different approach for model construction. In our approach rather than a priori assuming the relationship between the stimulus and the response from the cognisance of plant physiology we try and infer the relationship from the data (time course of response and stimulus traces). This approach is known as the black box modelling wherein the term black box in essence captures the opacity of mechanistic knowledge. An advantage of the black box modelling approach to the mechanistic counterpart is that we could also use the data to construct inverse model which could provide inference of the stimulus from the observed response. In our knowledge, this work is the first attempt to construct an inverse model of the stimulus driven electrical phenomenon in plants. Success of such inverse modelling would lead us towards developing environmental sensing applications using just the voltage recording from the surface of the plants.

In the present paper, we propose a novel dynamical model for describing the input-output characteristics of a bay leaf plant with change of light intensity following the system identification approach. In essence, the developed model predicts or detects the change of light intensity and the time duration for which light falls on the plant from measured time-varying electrical response data of the plant. A dynamical model is preferred over a static model since it takes into account the real-time changes of the internal state variables of a system expressible using one or few sets of ordinary differential equations involving temporal derivatives. Thus a dynamical model is considered more behavioural in nature than a static algebraic equation based model and have been preferred in our approach in this work. Since from the plant physiological point of view, the external light stimulus is responsible for the generation of the electrical response, the problem we address here is essentially an inverse problem where the characteristics of the light stimulus needs to be determined by observing the electrical response signal. Derivation of such inverse model is very challenging since there may exist a number of solutions for the same input-output behaviour. Similar studies on establishing an inverse relationship using linear and nonlinear system identification approaches for a real world system have been done in [29].

To address these issues, we applied two different classes of system identification techniques *viz*. Linear Time Invariant (LTI) discrete time estimators and nonlinear discrete time estimators. Then the most accurate model structure is chosen on the basis of obtained accuracies for estimating the system's parameter. Four different linear Least Square Estimator (LSE) based techniques, *viz*. Auto-Regressive eXogeneous (ARX), Auto-Regressive Moving Average eXogeneous (ARMAX), Box-Jenkins (BJ) and Output-Error (OE) [30] have been used within the LTI framework whereas, Non-Linear ARX (NLARX) and Non-Linear Hammerstein Winner (NLHW) [30] estimators have been considered within the nonlinear estimation framework. In this paper, we considered the input light stimulus as the input and the plant electrical response as the output for the forward modelling, thereby capturing the physical cause and effect relationship. Whereas with the inverse model, we tried to establish a causal relationship as a form of a dynamical model using electrical response as input and light stimulus as the output. The inverse model was proposed in order to make a prediction or detection framework to monitor the environment from plant electrical responses. To the best of our knowledge, this is the first study of its kind for establishing a forward and inverse causal relationship between the environment (light) and plant response. This gives a working solution for detecting the duration and nature of light inputs by only observing the electrical response. We have also tested the top three estimator structures, obtained from the rigorous exploration with light induced electrical signal for the *Laurus nobilis* plant on additional 19 different plants (17 *Zamioculcas zamiifolia* and two *Cucumis sativus* plants) under similar experimental conditions for the forward and inverse modelling, in order to show that these model structures give faithful prediction of the stimulus.

Rest of the paper is organized as follows. Section 2 provides a theoretical overview of system identification techniques describing the mathematical preliminaries of various estimators used in this paper. Section 3 briefly discusses the experiments for obtaining the plant data under light excitation and Section 4 describes the proposed model through comparison of the estimation accuracies between different estimators. Conclusions are drawn in section 5, followed by the references.

## 2. Theoretical background of forward/inverse dynamical system modelling and practical challenges

System modelling or identification can be viewed as a way of mathematically describing a phenomenon with some physical insight about the system from a measured input and output data-set. It is regarded as a bridge between the applications in real world problems and mathematical theories of model abstraction [30]. The application end of the spectrum of system modelling includes future prediction of the input to output characteristics of a model thus developed. A forward system model, where the response (output) needs to be related with the cause (the inputs), is much easier to develop in theory than developing an inverse model where the inputs are predicted from the observation of output response as this may result into one-to-many mapping situation. In essence, such a forward model will capture the dynamical characteristics of the response caused due to the excitation by establishing a physical cause-effect relationship. A similar approach by inverting the input and output may not always indicate physical causation but are capable of predicting the stimulus by only observing the response [29] which is adopted in the present study.

In principle, once formulated, a forward model could be inverted also to produce the inverse model from it. However, from the perspective of dynamical system theory, establishing a forward dynamical model (in terms of a transfer function) first and then inverting the poles and zeros may not always work if it contains non-minimum phase zeros (which gets converted to unstable poles of the inverse model) or possesses highly complex nonlinear terms which are explicitly noninvertible. Also in most cases of transfer function estimation for the forward problem using system identification approaches with any realistic data-set, the models are considered to have a proper transfer function structure i.e. the number of poles being higher than the number of zeros. This implies that the inverse model will be an improper transfer function with more zeros than poles leading to a high pass or differentiator type frequency response, thus leading to huge amplification of measurement noise in high frequency regions leading to a low signal to noise ratio (SNR). Therefore, a different approach may need to be adopted for solving the inverse modelling problem like the present one by inverting the recorded cause and effect, followed by varying the model structure between these two observed signals in order to match the cause in a best way. As has been mentioned in section 1, two major classes (and different sub-classes within each of them) of dynamic system parameter estimation techniques i.e. linear and nonlinear methods have been used in the current work for solving the inverse problem – which in our case, is predicting the characteristics of input light stimulus from the electrical response. In the following subsections, we briefly describe the theoretical backgrounds of each of these system identification techniques after describing the main statistical measures used here for measuring the accuracy of model fitting.

### 2.1. Least square estimation (LSE) technique for system identification

Considering the measured output and input of an unknown system at time $t$ is $y(t)$ and $u(t)$ respectively, the system can be described by a linear difference equation with co-efficient $a_i, i=1,\cdots,n$ and $b_j, j=1,\cdots,m$ as shown in equation (1)

$$y(t)+a_1 y(t-1)+\cdots+a_n y(t-n)=b_1 u(t-1)+\cdots+b_m u(t-m) \tag{1}$$

where, the estimated system parameter vector ($\theta$) and the measured input-output vector ($\varphi$) can be given as equation (2) and (3) respectively.

$$\theta = [a_1 \ldots a_n \ b_1 \ldots b_m]^T \tag{2}$$

$$\varphi(t) = [-y(t-1)\ldots -y(t-n) \ u(t-1)\ldots u(t-m)]^T \tag{3}$$

Using (2) and (3) the system model can be developed [6], incorporating a modelling error $e_t$ as represented by equation (4)

$$y_t = f(u_t, \theta) + e_t, \quad t=1,2,3,\ldots,N \tag{4}$$

We aim to find the certain parameter vector $\hat{\theta}$ which minimizes the least squared error ($S$) defined as

$$S \triangleq \sum_{t=1}^{N} e_t^2 = \sum_{t=1}^{N} (y_t - f(u_t, \theta))^2 \tag{5}$$

For practically useful case $f$ is linear in the unknown coefficients $\theta$. Hence equation (4) can be rewritten as equation (6) and further as equation (7) in matrix notation.

$$y_t = u_t^T \theta + e_t, \quad t = 1, 2, 3, \ldots, N \tag{6}$$

$$y = U\theta + e \tag{7}$$

Now the minimum error equation can be rewritten as equation (8)

$$S = e^T e = \left(y^T - \theta^T U^T\right)(y - U\theta) \tag{8}$$

The estimated value $\hat{\theta}$ which minimizes $S$ makes the gradient of $S$ with respect to $\theta$ as zero.

$$\frac{\partial S}{\partial \theta} = -2U^T y + 2U^T U\theta \tag{9}$$

Hence, the $\theta$ that makes the gradient of $S$ zero is given as

$$\hat{\theta} = \left[U^T U\right]^{-1} U^T y \tag{10}$$

Equation (10) describes the estimated system parameter for which the minimum of the sum of squared error over the time interval $1 \leq t \leq N$ is obtained and hence for this reason this method of estimation is called the least square estimation or LSE algorithm. It is worth noting that $\hat{\theta}$ is estimated using the measured input and output data and hence it can be easily inferred that using LSE, a parameterized system model can be developed. This technique hence is the backbone of all system identification methods.

## 2.2. *A linear system modelling using variants of LSE*

The accuracy and the efficiency of the system identification process depend on the choice of suitable estimator. In this section, a few variants of the estimators are discussed from the perspective of their applicability in system identification and hence the underlying choice of a certain estimator [31]. These estimators are derived from the consideration that the system can be described by a liner difference equation. A parameterized linear estimator in essence can be described by the following equation

$$y(t) = G(q^{-1}, \theta)u(t) + H(q^{-1}, \theta)e(t) \tag{11}$$

where, $y(t)$ and $u(t)$ are output and input to the system and $e(t)$ is a zero mean white Gaussian noise and $\theta$ is the parameter vector to be estimated, $G(q^{-1}, \theta)$ is the transfer function of the deterministic part (excitation to response) of the system and $H(q^{-1}, \theta)$ is the transfer function of the stochastic part (noise to response) of the system. Here $q^{-1}$ denotes the backward shift operator. Equation (11) can be further modified as equation (12) which is also known as the equation error type linear LSE.

$$A(q^{-1})y(t) = \frac{B(q^{-1})}{F(q^{-1})}u(t) + \frac{C(q^{-1})}{D(q^{-1})}e(t) \tag{12}$$

where, $\{B, F, C, D\}$ are polynomials in $q^{-1}$ and represent the numerator and denominator of the system and noise model respectively and $\{A\}$ represents the polynomial containing common set of poles for both the system and noise model. The block diagram representation of the generalized equation error type LSE in equation (12) is shown in Figure 1(a).

The generalized LSE can be further customized by considering fewer combinations of the polynomials $\{B, F, C, D\}$ at once, which, in its essence, paves the path towards the choice of a suitable linear estimator for system identification. An example of such a customization is the FIR filter form of the generalized LSE while considering only polynomial $\{B\}$. In the following subsections, a brief description of four such classes of estimators as few variants of generalized equation error type LSE are furnished.

### 2.2.1. Auto Regressive eXogenous (ARX) estimator

The basic structure of the ARX estimator is governed by equation (13).

$$A(q^{-1})y(t) = B(q^{-1})u(t) + e(t) \tag{13}$$

The main disadvantage of this structure is that the deterministic (system) and the stochastic (noise) dynamics are both modelled with the same set of poles, which may be unrealistic in many applications.

### 2.2.2. Auto Regressive Moving Average eXogenous (ARMAX) estimator

The basic structure of the ARMAX estimator is governed by equation (14).

$$A(q^{-1})y(t) = B(q^{-1})u(t) + C(q^{-1})e(t) \tag{14}$$

The major advantage of the ARMAX structure over the ARX structure is that it suffices with a better flexibility to model the exogenous noise dynamics by introducing a moving average to the white noise. Although the ARMAX estimates same set of poles, it estimates different set of zeroes for both the system and the noise components. Thus ARMAX is useful where the entire system dynamics is dominated by the stochastic dynamics (noise).

### 2.2.3. Box-Jenkins (BJ) estimator

The basic structure of the BJ estimator is governed by the following equation (15).

$$y(t) = \frac{B(q^{-1})}{F(q^{-1})} u(t) + \frac{C(q^{-1})}{D(q^{-1})} e(t) \tag{15}$$

BJ structure allows the estimation of different set of poles and zeroes for the system and noise component. This structure is especially useful when noise enters the system at a later stage e.g. measurement noise.

### 2.2.4. Output-Error (OE) estimator

The estimator has the following structure (16).

$$y(t) = \frac{B(q^{-1})}{F(q^{-1})} u(t) + e(t) \tag{16}$$

The OE structure estimates the poles and the zeroes of the system model only and estimation of the noise model is ignored. This structure can be used when the deterministic dynamics dominates the overall system dynamics and the stochastic dynamics has no significant effect.

### 2.3. *Estimation of nonlinear models*

For any nonlinear dynamical system whose input and output are $u(t)$ and $y(t)$ respectively, can be expressed as equation (17) [32].

$$y(t) = f\left(u(t-1), y(t-1), u(t-2), y(t-2), \cdots\right) \quad (17)$$

where, $f(\cdot)$ is a nonlinear function representing any arbitrary nonlinearity. The nonlinear black-box identification is done using two nonlinear models and the parameters are found out by estimation of squared error as discussed in the previous section. The nonlinear estimators are *Nonlinear-ARX* and *Nonlinear-Hammerstein-Wiener* having the nonlinearity in parallel and series connection respectively with the basic linear blocks. Brief descriptions of these two classes of estimators are given in the following subsections.

#### 2.3.1. Nonlinear ARX (NLARX) estimator

In principle the NLARX is an extension to the linear ARX estimator. A linear ARX model can be expressed as equation (18) [32]

$$y(t) + a_1 y(t-1) + a_2 y(t-2) + a_n y(t-n) = b_1 u(t) + b_2 u(t-1) + b_m u(t-m) + e(t) \quad (18)$$

Equation (13) describes the linear ARX structure which implies that the current output is predicted as a weighted sum of past output values and also current and past input values. Rewriting equation (18) as a product form equation (19) is obtained.

$$y_p(t) = [a_1, a_2, ..., a_n, b_1, b_2, ..., b_n] \begin{bmatrix} y(t-1), y(t-2), ..., y(t-n), \\ u(t), u(t-1), ..., u(t-m) \end{bmatrix}^T \quad (19)$$

where, $y_p(t)$ is the current output and $\{y(t-1), \cdots, u(t-1), \cdots\}$ are delayed input and output variables called the regressors. Instead of weighted sum as described in equation (19) the predicted output $y_p(t)$ can also be mapped using a nonlinear mapping function $f(\cdot)$ which gives rise to the structure NLARX as described in equation (17). The NLARX is composed of a nonlinear estimator or simply a nonlinear mapping function part and a regressor part which in turn is the collection of delayed input and output variables. The block diagram representation of the NLARX model is shown in Figure 1(b) [32]. The nonlinear estimator can be further composed of a parallel connection of a linear and a nonlinear function block. The nonlinearity estimator block maps the regressors to the model output using a combination of nonlinear and linear functions. Various nonlinearity estimators, such as tree-partition networks, wavelet networks, and neural networks can be selected for both input and output blocks.

#### 2.3.2. Nonlinear Hammerstein-Wiener estimator

For certain systems when the output of the system depends nonlinearly with its input, the input-output relationship can be broken down to a series of interconnected elements i.e. the system dynamics is represented by using a linear transfer function and the nonlinearities are captured using nonlinear mappings of the input and output [32]. In Hammerstein-Wiener estimator, such an approach has been adopted where a linear block is connected in series with two static nonlinearities. The block diagram representation of the Hammerstein-Wiener structure is shown in Figure 1(c) [32]. Here, the predicted output $y_p(t)$ is as follows:

$$y_p(t) = h\left(\frac{B(q^{-1})}{F(q^{-1})}(f(u(t)))\right) \quad (20)$$

where, $f(t)$ is a nonlinear function transforming the input data. $\frac{B(q^{-1})}{F(q^{-1})}$ is a linear transfer function with $\{B,F\}$ being rational polynomials similar to the output error model and $h(t)$ is another nonlinear function that maps the output of the linear block to the system output.

Here, the input nonlinearity is a static function, implying that the output at a given time depends only on the input at that time. The input nonlinearity can be varied as a sigmoid network, wavelet network, saturation, dead-zone, piecewise linear function, one-dimensional polynomial, or some user-defined custom network [32]. The input or output nonlinearity can also be not used at all. The linear block may be configured by specifying the orders of the numerator $\{B\}$ and denominator $\{F\}$. Just like the input nonlinearity is static, the output nonlinearity is a static function as well. The output nonlinearity may be configured in the same way as the input nonlinearity. In all simulations presented in this paper, the Levenberg-Marquardt search method has been used to optimize the parameters of nonlinear models with a criterion for minimizing the determinant of squared error ($\det(e^T e)$).

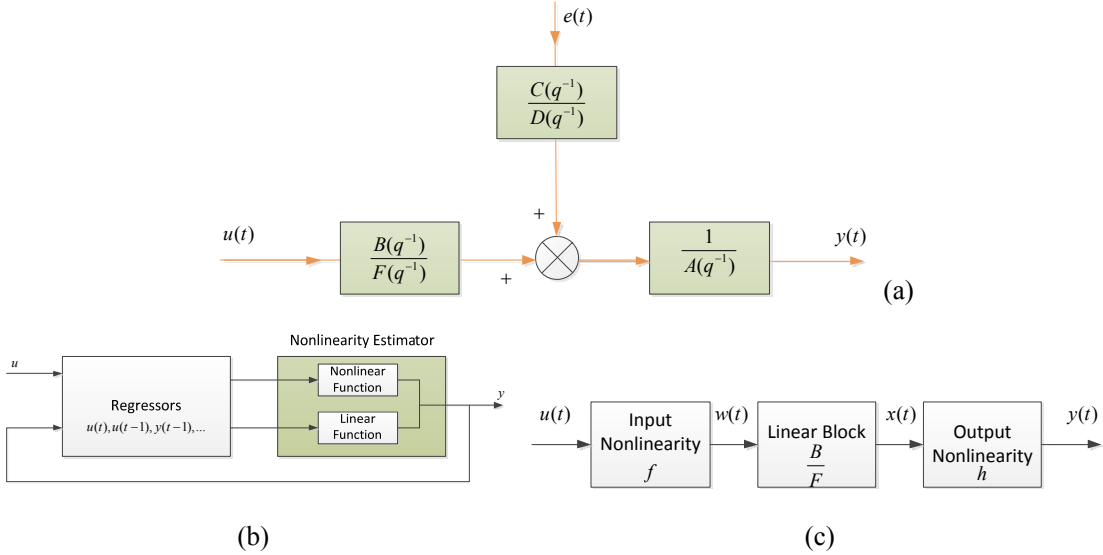

Figure 1: Block diagram representation of the linear and nonlinear estimators (a) linear generalized least-square estimator (b) nonlinear ARX estimator (c) nonlinear Hammerstein-Wiener estimator.

## 3. Experimental data and model estimation

This section gives a brief description about the experimental set-up used for the collection of electro-physiological signals of one bay leaf (*Laurus nobilis*), two cucumber (*Cucumis sativus*) and 17 Zanzibar Gem (*Zamioculcas zamiifolia*) plants. These plants were subjected to a periodic white light stimulus of different pulse widths. Approximately, 1 week to 2 years old pot-grown plants of different species were chosen for the experiments. The measurement of electrical signal responses of these plants were done by inserting two thin metallic Electromyogram (EMG) needle electrodes (Bionen s.a.s) into the petiole and stem at a distance of approximately 5 cm from each other. The third reference electrode was inserted into the plant body, nearer to the soil. A dual instrumentation amplifier (EI-1040) [33] with a gain of $10^3$ was chosen to provide the high input impedance without altering the actual amplitude of the acquired signal. For low frequency and low amplitude signals, we require very high input impedance [34] and very low input bias currents. The EI-1040 provides an input impedance of 10 GΩ and an input bias current of 0.5 nA. The differential output of the amplifier is provided in input to the USB6008 [35], a National Instruments Data Acquisition (DAQ) board featuring 8 analog inputs (12-bit, 10 kS/s). The USB6008 was used for analog to digital (A/D) conversion at a sampling rate of 1 kHz which was then monitored using LabVIEW 2012 software [36] on a personal computer (PC). Since bio-electrical signals are usually weaker [37], interference from an external electromagnetic field could induce a lot of noise in it. Therefore this setup (excluding the PC) was placed inside a grounded Faraday cage. The LabVIEW software also controls an Arduino microcontroller [38] that pilots the emission of light possibly according to some specific patterns (e.g. 5 minutes of light, 10 minutes of dark). The following schematic diagram in Figure 2(a) shows the connection of the experimental set-up used to acquire the data on the signals generated by plants in reaction to light stimuli.

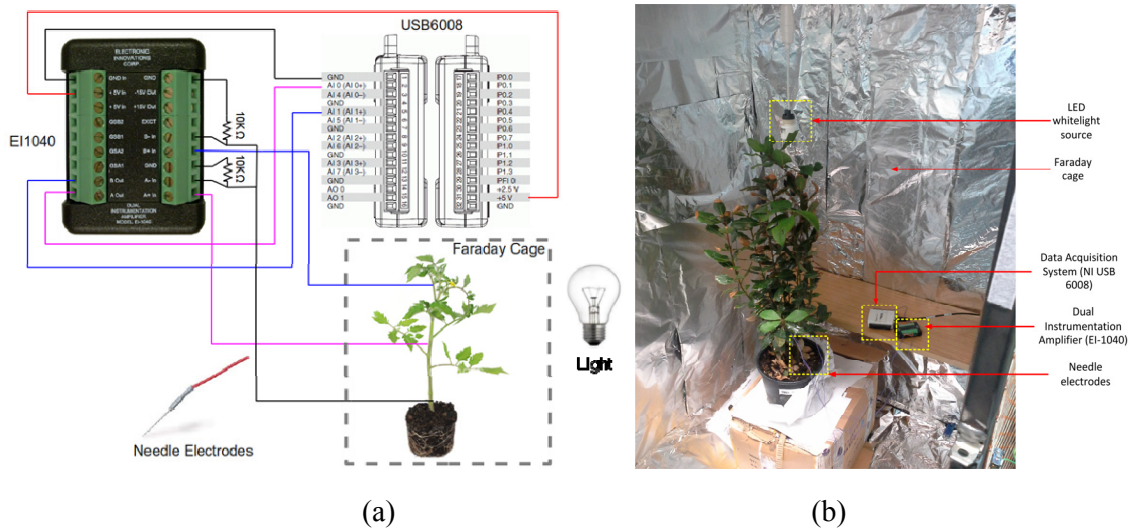

(a) (b)

Figure 2: (a) Schematics of the connection between the different devices employed to capture the electrical signals generated by plants in reaction to light stimuli. (b) Experimental setup for a bay leaf plant with white-light excitation

A digital low-pass filter with a cut-off frequency 1 Hz was provided to eliminate any noise associated with the measurement, as in [13–18], [39] it is reported that plant signals are slow oscillatory signals at a very low frequency. Light Emitting Diode (LED) light source were used for providing white-light at maximum brightness. Here, the number of photons used for photosynthesis by plants is used as basis for measurement of incident light (in Photo-synthetically active radiation or PAR unit). Similar experiment and electrophysiological measurements on cucumber plants can be found in [40]. In order to get a robust model, the light pulse widths were varied for 20 different plants during the experiment and are given in Table 1.

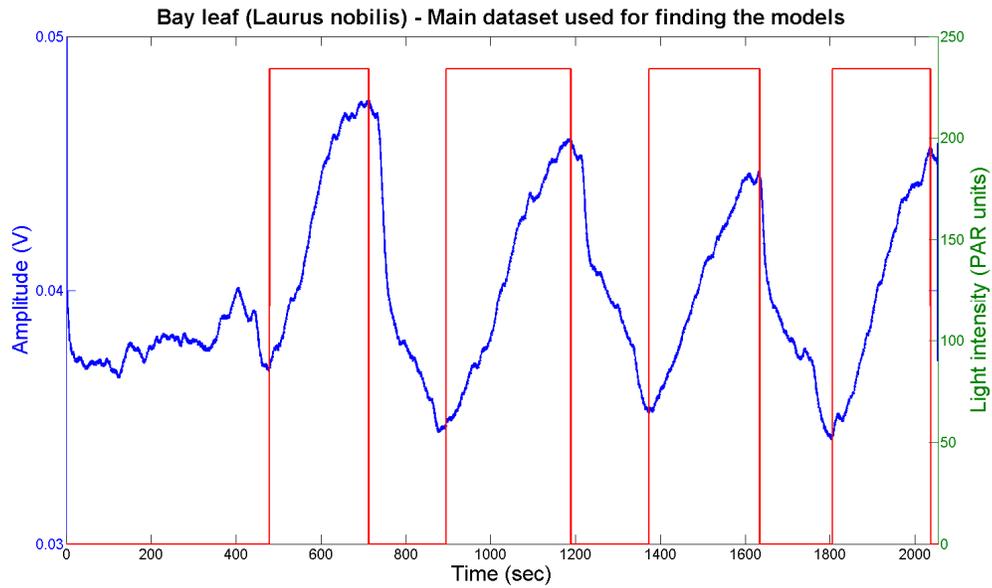

(a)

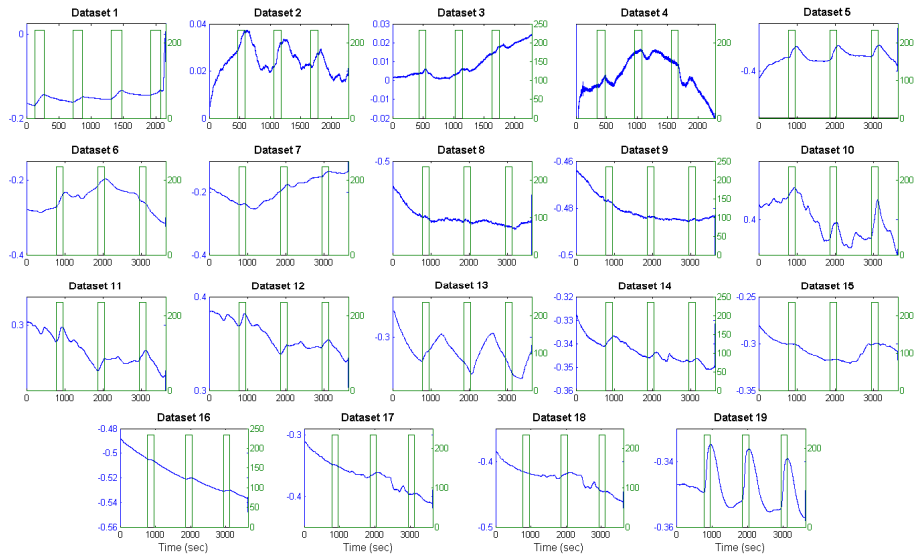

(b)

Figure 3: Plot of the variations in electrical signal response of the plant with respect to the incident light stimulus (a) main dataset (b) 19 independent test datasets.

Figure 2(b) shows the experimental setup used to obtain data for estimating the forward and inverse models. Aluminium foil was used to cover up the inside of the faraday cage from all sides to shield radio-frequency (RF) interference. The LED bulb was hung from the top to focus on the intended portion of the bay leaf plant. The electrodes were connected to the instrumentation amplifier whose output was connected to the DAQ. The DAQ was connected to the PC, sitting outside the faraday cage, through a USB cable.

Figure 3(a) shows the electrical signal responses of the main dataset when exposed to light pulses of varying widths. The y-axis of Figure 3(a) shows both the amplitude of the electrical responses and the light intensities in two different axes. Also, 19 independent test datasets are used in Figure 3(b) to verify whether a common modelling framework is capable of successfully predicting the stimulus for different plants as well. It is observed in all the cases that whenever the light pulse is switched on or off, there is a clear change in the gradient of the electrical response of the plant which we are trying to model in the present work. These electrical responses are induced on the leaf tissue of the plants. However it is worth noting that extracellular measurements of the membrane potential on a plant leaf is a mixture of the individual responses of the guard cells, mesophyll cells and the epidermal cell [15]. The morphology of the membrane potential of these cells can be different from what have been recorded on the surface of the leaf tissue. We intent to arrive at a black box model and thus the collective morphology of the membrane potential found on the tissue will suffice our purpose of model building.

Table 1: Variations in white-light pulse widths for each dataset (for twenty different plants)

| Datasets | Plant species | Time (sec) | | | |
|---|---|---|---|---|---|
| | | First light pulse | Second light pulse | Third light pulse | Fourth light pulse |
| Main | *Laurus nobilis* | 232.401 | 293.201 | 260.001 | 231.001 |
| Dataset 1 | *Zamioculcas zamiifolia* | 147.801 | 151.801 | 167.001 | 148.001 |
| Dataset 2 | *Zamioculcas zamiifolia* | 123.2 | 120 | 120 | - |
| Dataset 3 | *Zamioculcas zamiifolia* | 119.999 | 119 | 121 | - |
| Dataset 4 | *Zamioculcas zamiifolia* | 140.998 | 119.998 | 121 | - |
| Dataset 5,6,10-19 | *Zamioculcas zamiifolia* | 180 | 180 | 180 | - |
| Dataset 7,8 | *Cucumis sativus* | 180 | 180 | 180 | - |

### 4. Forward and inverse modelling results using system identification technique

We now proceed with a notion that the electrical signals generated from plants has come from a black box system without the need of having any prior knowledge of its internal electrophysiological dynamics. With the measured stimulus (light) and response (electrical signal) data-set, dynamical models are developed using the concept discussed in section 2. For the present simulation studies, we used the System Identification Toolbox of MATLAB [32] to develop the input-output linear and nonlinear forward/inverse models. The idea is to develop a model whose predicted output best fits the experimentally recorded or measured output when the same input is applied.

As explained earlier, the inverse modelling is incorporated to simulate the switching instants of the stimulus, treating the actual electrical response of the plant as the input to the model and treating the actual photon flux density of the light, incident on the plant as an output from the model. The goal was to find the error between the rise and fall time of the measured and predicted light pulses. The error between the peak values of measured and predicted amplitudes of the photon flux densities of the light pulse was also compared for different light pulses in the subsequent simulations. Various estimator structures from the linear and nonlinear system identification techniques are used to model the electrical response of the plants. Since for the linear estimators, the obtained transfer functions represent discrete time models in $z$-domain ($z$ being the discrete time complex frequency), they can easily be converted to continuous time transfer function models (i.e. in $s$-domain with $s$ being the Laplace variable or continuous time complex frequency) by using the well-known bilinear transform (21) involving the sampling time $T_s$.

$$s = \frac{2}{T_s}\left(\frac{1-z^{-1}}{1+z^{-1}}\right) \tag{21}$$

The percentage fit as shown in the figures to compare relative accuracies of different estimators is given by the normalized root mean squared error (NRMSE) as given in (22).

$$fit = \left[1 - \frac{\|y - \hat{y}\|}{\|y - \bar{y}\|}\right] \times 100 \; \% \tag{22}$$

where, $\hat{y}$ is the simulated or predicted model output, $y$ being the measured output and $\bar{y}$ is the mean of the output.

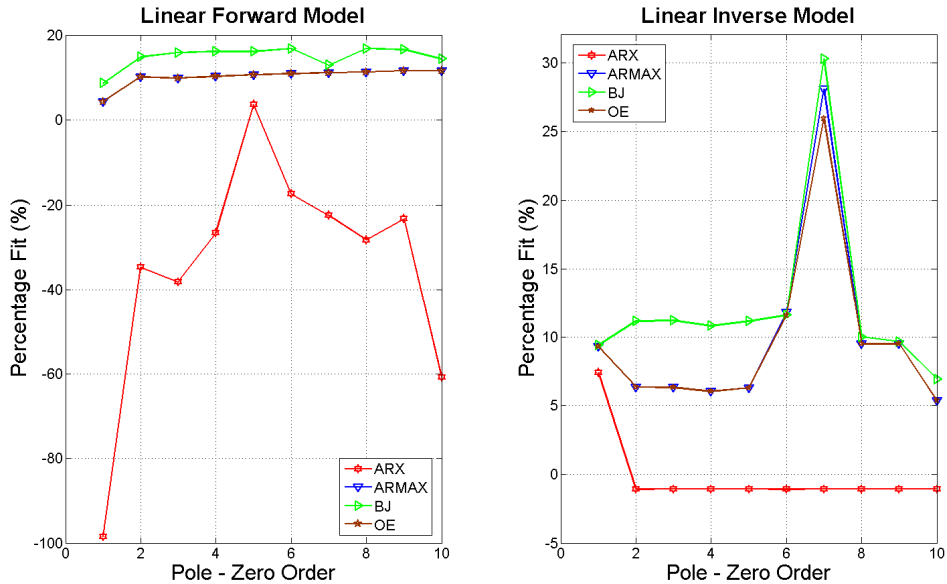

Figure 4: Linear models for forward and inverse modelling using the main dataset.

In Figure 4, the plots having the best fits using linear models for both forward and inverse scenario of the main dataset are shown while the y-axes shows the respective percentage fits with gradual increase in pole-zero order for the linear models. All the four variants of linear LSE i.e. ARX, ARMAX, BJ, OE are compared indicating that BJ model outperforms the others in both cases as it is consistently giving higher percentage accuracy than that with other variants.

It can also be observed from Figure 4 that in forward modelling, the ARX model shows a high variation (or less consistency) in terms of percentage fit as the order is varied. Whereas, ARMAX, BJ and OE models are consistent while predicting the output electrical signal, when the light pulse is fed as the model input. When observing the case of inverse modelling, ARX model is more consistent than the other three models with gradual increase in pole-zero order. The other three models show a sudden jump in accuracy, predicted in terms of percentage fit, at a combination of pole-zero order as 7 (i.e. 7 poles, 7 zeros, and one sample delay).

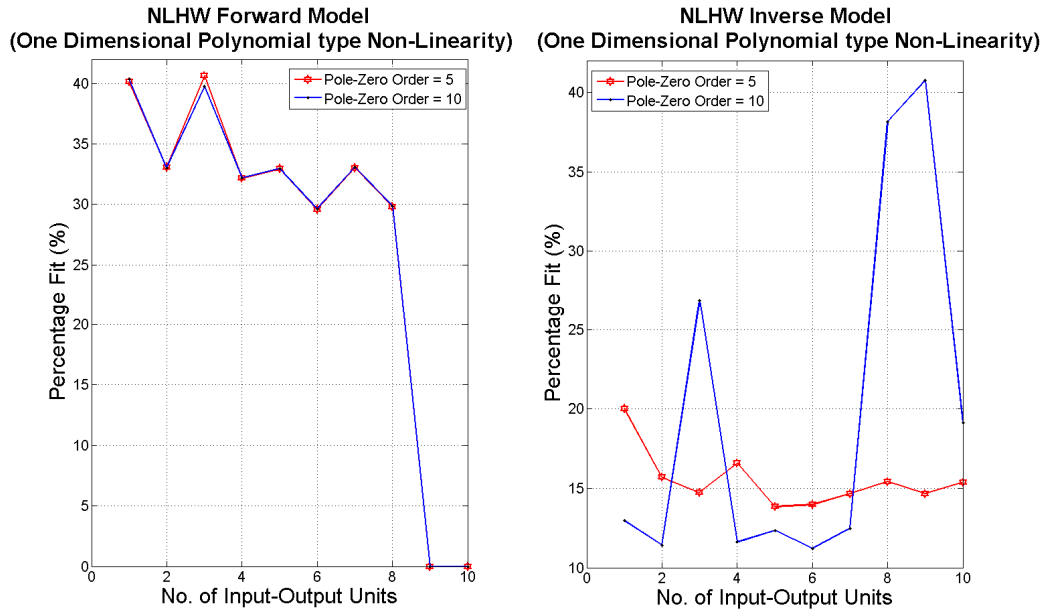

Figure 5: NLHW model with one dimensional polynomial as nonlinearity for forward and inverse modelling using the main dataset.

We now proceed towards similar simulations with nonlinear estimators like few NLARX variants e.g. wavelet network, sigmoid network and tree-partition, with gradual increase in the number of regressors for input and output signal in both forward and inverse modelling scenario. But it consistently resulted in negative percentage fit and thus has not been reported in the paper. On contrary, the NLHW estimators (with different static input-output nonlinearity) in most cases yielded good prediction accuracy which is compared in the following simulations.

Figure 5 shows the NLHW model with one dimensional polynomial as the nonlinearity type. In this case, we have chosen two pole-zero orders, 5 and 10 and varied the number of input-output units for each order. While considering the forward model, we see a lot of variance in percentage fit between input-output units from 1 through 4. Thereafter, the percentage fit seems to stabilize till input-output units till 8 and the fit suddenly drops beyond 8 input-output units. In the inverse modelling case, we see a less variability in prediction in terms of percentage fit using pole-zero order of 5 than polynomial order 10.

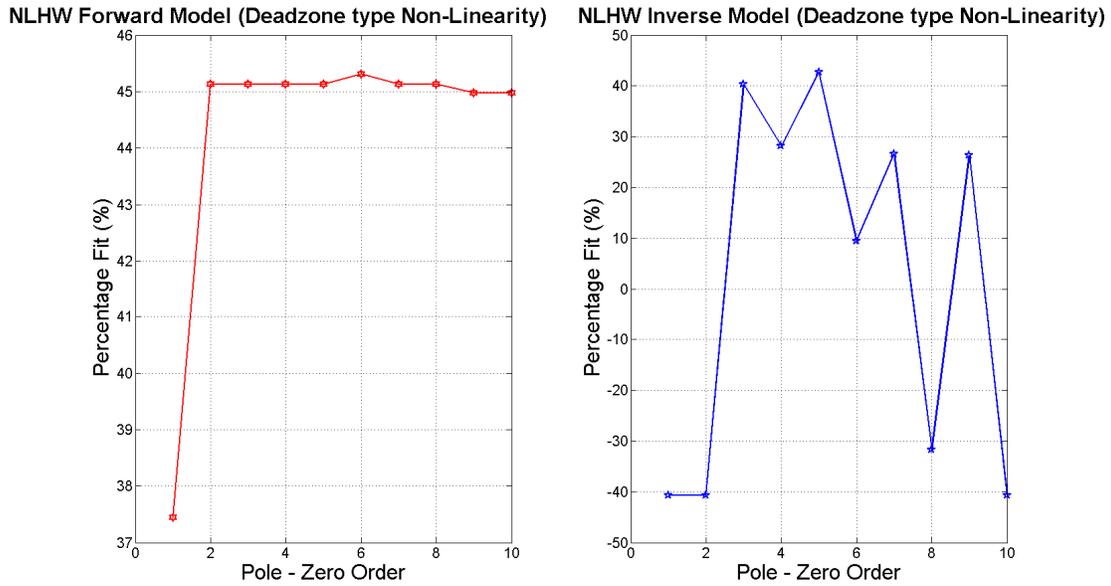

Figure 6: Nonlinear model with dead-zone as nonlinearity for forward and inverse modelling using the main dataset.

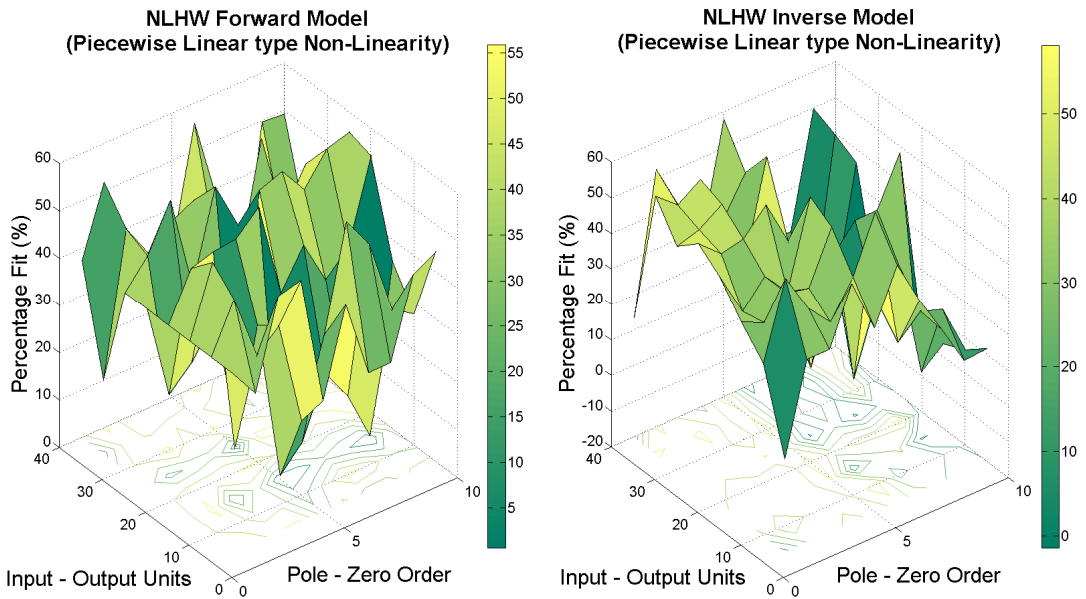

Figure 7: Nonlinear model with piecewise linear as nonlinearity for forward and inverse modelling using the main dataset.

While using the dead-zone type static nonlinearity as shown in Figure 6, we observe a stable prediction during forward modelling for pole-zero orders higher than two. However for inverse modelling with the same estimator configuration, it has been found that the variation in the percentage fit is quite high.

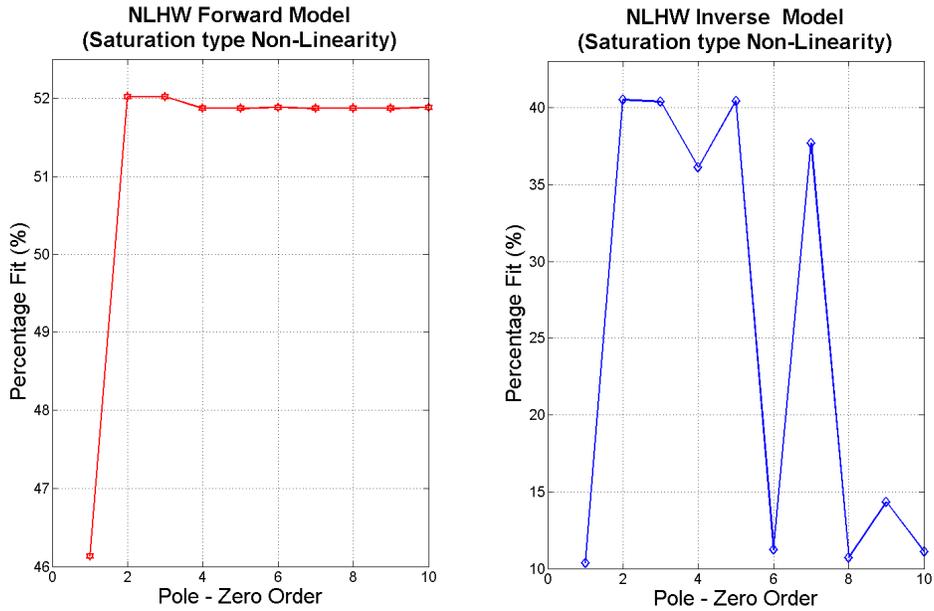

Figure 8: Nonlinear model with saturation as nonlinearity, for forward and inverse modelling using the main dataset.

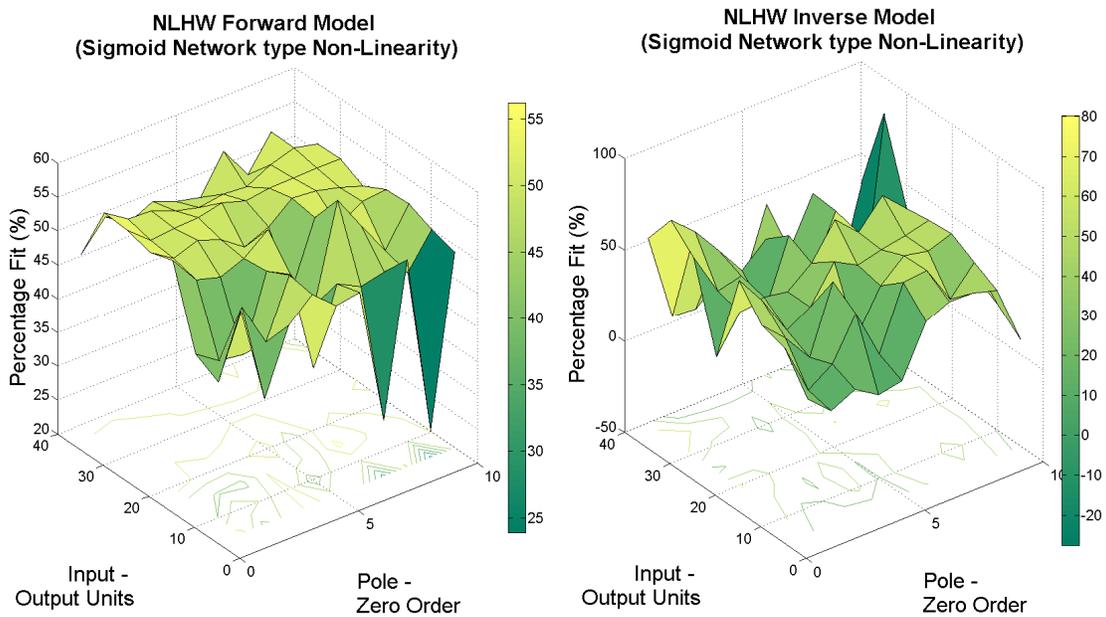

Figure 9: Nonlinear model with sigmoid as nonlinearity, for forward and inverse modelling using the main dataset.

For piecewise linear type static nonlinearity in NLHW estimator, we varied both the input-output units and also the pole-zero orders for each case. Thus we obtain a surface plot of percentage fit as a function of these two which has been shown in Figure 7. Here the input output units were varied from 5 through 40, incrementing by 5 at a time. For each chosen pair of input-output units, the pole zero order were varied from 1 through 10. When considering forward modelling, we often see a percentage fit around 50% using input-output units of 20 onwards and pole-zero order between 2 and 8. In the case of inverse modelling, we can see a percentage fit of 50% when the pole-zero order are between 1 and 6.

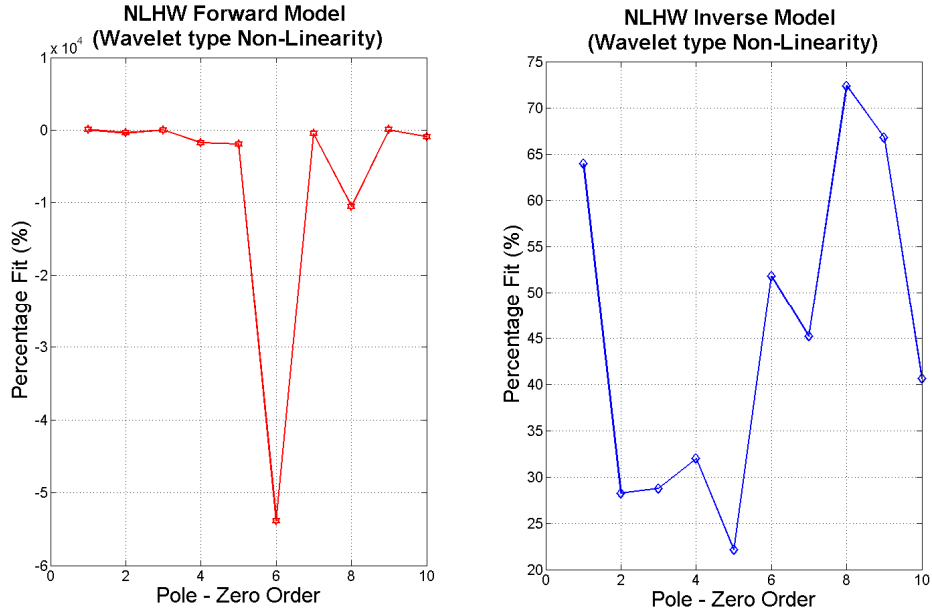

Figure 10: Nonlinear model with wavelet network as nonlinearity, for forward and inverse modelling using the main dataset.

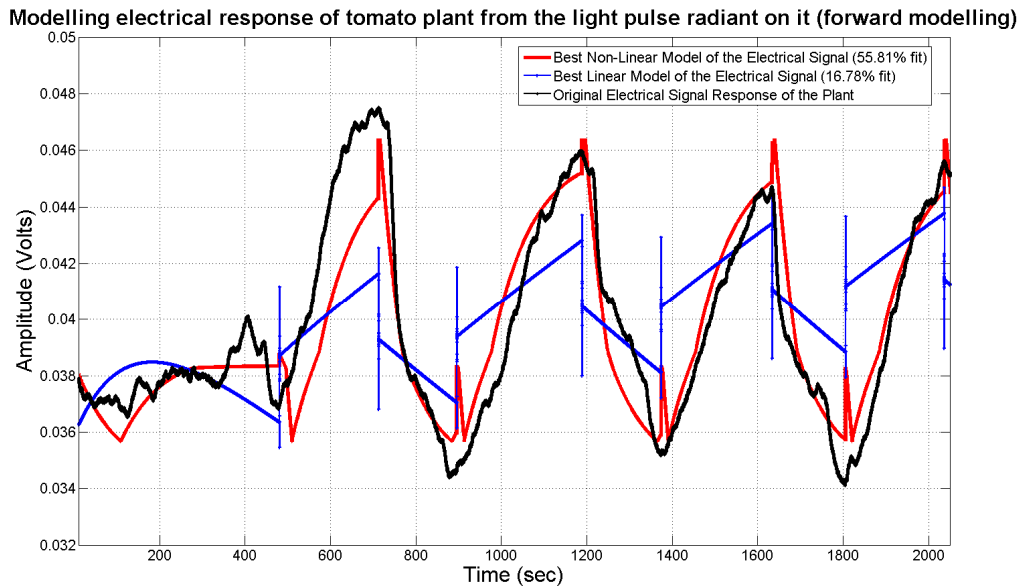

Figure 11: Best linear and nonlinear model estimates during forward modelling using the main dataset.

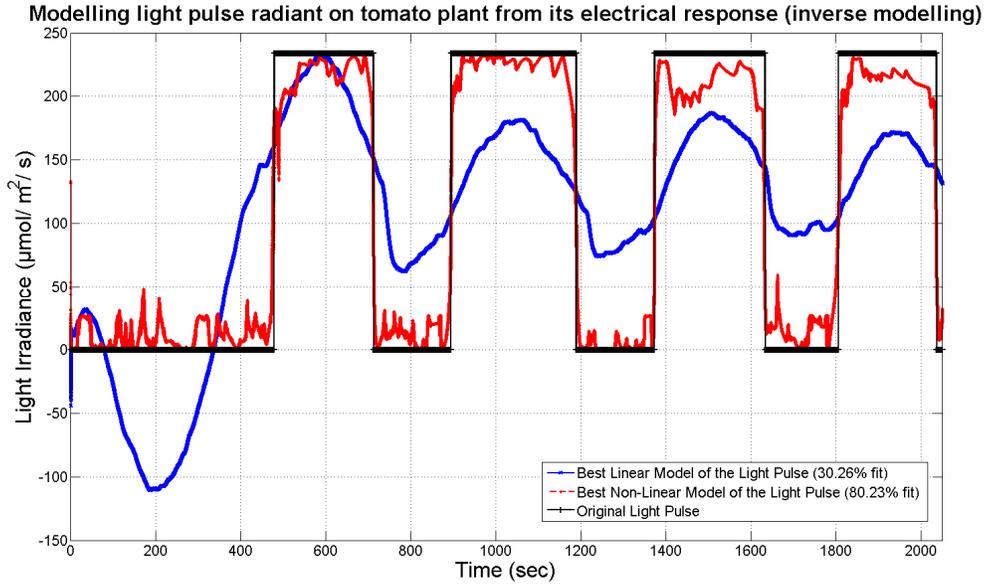

Figure 12: Best linear and nonlinear model estimates during inverse modelling using the main dataset.

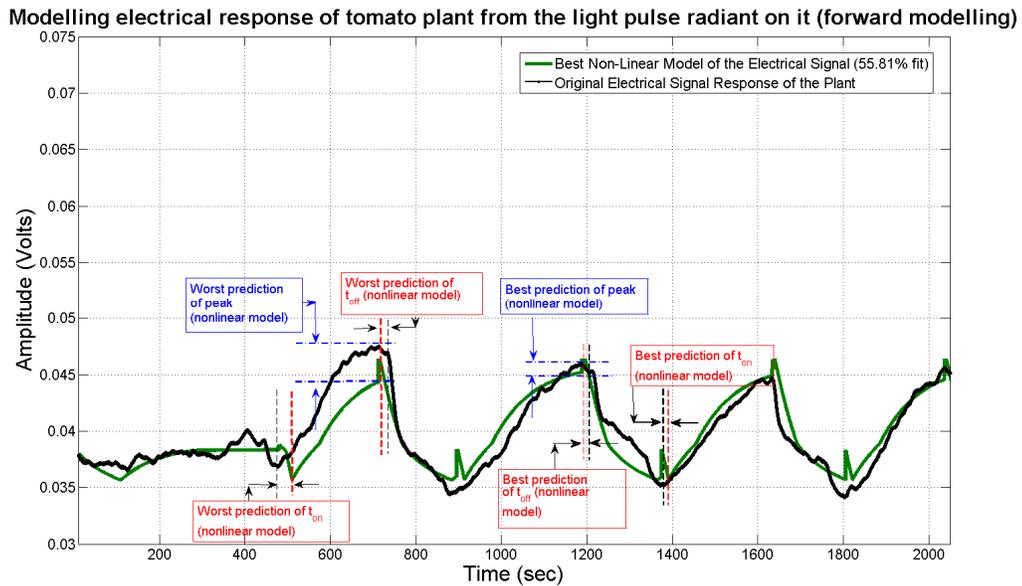

Figure 13: Measuring best and worst prediction of peaks and rise times and fall times of the plant electrical signal (forward modelling) using the main dataset.

In Figure 8, we can see that using saturation type nonlinearity in forward modelling gives consistent prediction accuracy for pole-zero order above 1. For the inverse modelling case, some sort of consistency in accuracy is shown for pole-zero order between 2 and 5 and randomly varying otherwise. When using sigmoid type static nonlinearity, we can see from Figure 9 that for forward modelling, percentage fit peak has occurred at around 50% when input-output unit is 30 and then decreases thereafter. For inverse modelling, a peak fit of 80% is noticed for number of units of 35 and pole-zero order of 10.

Table 2: Best forward model with % fit, $t_{on}$, $t_{off}$ and peak while tested on the main dataset

| Estimator configuration | % fit | Best Case | | | Worst Case | | |
|---|---|---|---|---|---|---|---|
| | | $t_{on}$ (sec) | $t_{off}$ (sec) | peak (Volts) | $t_{on}$ (sec) | $t_{off}$ (sec) | peak (Volts) |
| Piecewise linear-10,10,1 (nonlinear) | 55.81 | 7 | 4 | 0.00079 | 0.4 | 2 | 0.00431 |
| BJ-6,6,6,6,1 (linear) | 16.78 | 9.3 | 1 | 0.00111 | 7.2 | 6 | 0.00608 |

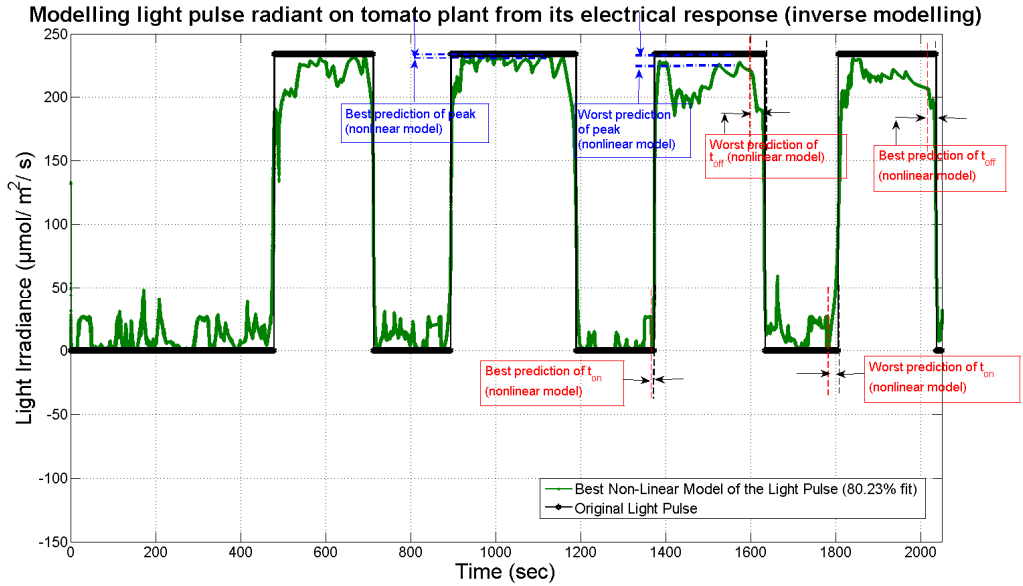

Figure 14: Measuring best and worst prediction of peaks and $t_{on}$, $t_{off}$ of the predicted light-pulse (inverse modelling) using the main dataset.

Table 3: Best inverse model with %fit, $t_{on}$, $t_{off}$ and peak while tested on the main dataset

| Estimator configuration | % fit | Best Case | | | Worst Case | | |
|---|---|---|---|---|---|---|---|
| | | $t_{on}$ (sec) | $t_{off}$ (sec) | peak (µmol/m²/sec) | $t_{on}$ (sec) | $t_{off}$ (sec) | peak (µmol/m²/sec) |
| Sigmoid-10,10,1 (nonlinear) | 80.23 | 7.1 | 5.2 | 2.0 | 9.5 | 6.0 | 45 |
| BJ-7,7,7,7,1 (linear) | 30.26 | 13 | 26 | 6.7 | 18.8 | 51.2 | 63 |

Lastly, when we consider wavelet network type of static nonlinearity (radial function wavelet) in Figure 10, during forward modelling we notice fairly consistent prediction accuracy except pole-zero order of 6 and 8. During inverse modelling we observe a high variability in the prediction accuracy with gradual increase in pole-zero order.

We now compare the best found forward models from different variants of linear and nonlinear estimators tested on the main dataset in Figure 11 showing the predicted signals using the NLHW model with piecewise linear type nonlinearity (10 pole-zero order and 25 input-output units) and linear BJ model with 6 pole-zero order. For the inverse model the best configurations are NLHW with sigmoid nonlinearity (10 pole-zero order and 35 input-output units) amongst the nonlinear estimators and BJ with 7 pole-zero order amongst the linear estimators which has been shown in Figure 12. From the best inverse models the predicted light pulse widths are captured in terms of the on-time ($t_{on}$) and off-time ($t_{off}$).

Table 4: Top three estimator settings for the main dataset during forward and inverse modelling

| Class of models | Model number | Nonlinearity in NLHW estimator | Input Units | Output Units | Poles | Zeroes | Delay |
|---|---|---|---|---|---|---|---|
| Inverse Models | 1 | Sigmoid | 35 | 35 | 10 | 10 | 1 |
| | 2 | Wavelet | - | - | 8 | 8 | 1 |
| | 3 | Sigmoid | 35 | 35 | 1 | 1 | 1 |
| Forward Models | 1 | Piecewise Linear | 25 | 25 | 10 | 10 | 1 |
| | 2 | Piecewise Linear | 40 | 40 | 6 | 6 | 1 |
| | 3 | Piecewise Linear | 25 | 25 | 9 | 9 | 1 |

Next, we calculate the best and worst prediction in terms of on-time and off-time for the light pulse (measured versus predicted in the inverse modelling case) or rise time and fall time for the electrical signal (measured versus predicted in the forward modelling case). We also note down the difference between the measured and predicted peaks of light intensity (inverse) and electrical signal (forward). The methodology for defining the best and worst prediction accuracy for the $t_{on}$, $t_{off}$ and peak amongst different light pulses are shown in Figure 13 and Figure 14 respectively for the forward and inverse modelling perspective with the best and worst values of these three parameters reported in Table 2 and Table 3 respectively. It is evident that the linear model outputs show low percentage fit in terms of prediction accuracy when compared to the nonlinear model.

A positive value on $t_{on}$ or $t_{off}$ denotes the simulated result occurs before the actual rise and fall time of the pulse. In some cases using NLARX estimator it yield negative percentage fit. In such cases a negative value of $t_{on}$ or $t_{off}$ denotes the simulated value occurs after the actual event. Similarly, a positive value of the predicted peak amplitude denotes a higher than actual value while a negative value denotes a lower than actual value. Figure 11 and Figure 12 shows the best linear and nonlinear model estimates for forward and inverse modelling cases. The parameters mentioned in Table 2 and Table 3 for Box-Jenkins model is $n_b, n_f, n_c, n_d$ and $n_k$ denoting the order of the polynomials for the deterministic and stochastic part of the model and the delay unit respectively as discussed in section 2.2.3. Hence the term BJ-7,7,7,7,1 (or BJ-6,6,6,6,1) in Table 2 and Table 3 implies that the values of the parameters respectively. Similarly for NLHW class of models with sigmoid and piecewise linear type static nonlinearity, the number of poles, zeroes and delay-unit is given by (10,10,1) respectively.

The main dataset (on *Laurus nobilis* plant) was used for rigorous parameter estimation of different model structures. We then tested the top three estimator settings (termed as model 1, 2 and 3) for both the forward and inverse problem which yielded the best accuracies (as shown in Table 4). Thereafter, we applied the estimator configurations on 19 other independent datasets (17 *Zamioculcas zamiifolia* and two *Cucumis sativus* plants) to validate if they consistently produce acceptable forward and inverse prediction which have been reported in Table 5. It is evident from Table 5 that even in other datasets, the NLHW models give positive prediction accuracy. Also, the inverse models produce comparable accuracy under the same configuration. Although in Figure 3(a) and (b), there are slightly different morphology of electrical signals in different plants, the forward and inverse estimation accuracies seems promising. The best setting of the estimators are found from the main dataset and hence some deviation or fall in prediction accuracy by the same estimator configuration for the other data-sets are expected.

Table 5: Performance of the top three estimator structures (in %) for each of the 20 datasets during forward and inverse modelling

| Dataset | Temperature (°C) | Humidity (%) | Inverse Models (% fit) | | | Forward Models (% fit) | | |
|---|---|---|---|---|---|---|---|---|
| | | | Model-1 | Model-2 | Model-3 | Model-1 | Model-2 | Model-3 |
| Main | 23.8 | 49 | 80.23 | 72.36 | 69.53 | 55.81 | 55.76 | 54.42 |
| 1 | 24.4 | 55 | 59.70 | 5.23 | 7.72 | 26.09 | 0.62 | 25.72 |
| 2 | 24.2 | 57 | 64.88 | 18.45 | 7.58 | 40.81 | 40.46 | 40.32 |
| 3 | 23.7 | 52 | 38.04 | 31.32 | 17.80 | 85.49 | 94.97 | 83.86 |
| 4 | 24.6 | 49 | -2.91 | 26.34 | 36.49 | 14.01 | 62.98 | 19.50 |
| 5 | 22.8 | 47 | 45.29 | 48.06 | 63.90 | 59.24 | 33.43 | 44.14 |
| 6 | 22.4 | 54 | 2.07 | 12.73 | 62.17 | 27 | 70.44 | 14.28 |
| 7 | 26.3 | 47 | 56.38 | 28.15 | 55.81 | 91.88 | 42.43 | 91.88 |
| 8 | 23.3 | 49 | 28.04 | 54.76 | 53.46 | 21.50 | 50.28 | 78.57 |
| 9 | 23.3 | 49 | 22.01 | 73.90 | 8.75 | 86.86 | 91.85 | 85.74 |
| 10 | 24.4 | 53 | 36.39 | 53.41 | 54.12 | 42.09 | 65.40 | 20.40 |
| 11 | 24.6 | 52 | 2.23 | 9.22 | 56.50 | 67.81 | 66.27 | 65.14 |
| 12 | 24.6 | 52 | 5.64 | 53.41 | 3.18 | 67.76 | 26.05 | 66.03 |
| 13 | 23.7 | 52 | 69.51 | 17.78 | 41.79 | 50.44 | 39.10 | 18.81 |
| 14 | 24.8 | 56 | 72.99 | 26.46 | 34.30 | 23.56 | 77.90 | 79.38 |
| 15 | 23.1 | 50 | 25.34 | 60.90 | 47.93 | 90.41 | 91.81 | 93.31 |
| 16 | 23.1 | 35 | 30.42 | 75.06 | 67.38 | 97.70 | 98.34 | 46.49 |
| 17 | 23.1 | 50 | 31.30 | 71.77 | 53.55 | 91.18 | 11 | 90.48 |
| 18 | 23.1 | 50 | 72.95 | 31.45 | 54.02 | 74.95 | 18.77 | 25.08 |
| 19 | 24.9 | 58 | 50.43 | 30.44 | 67.81 | 72.20 | 57.42 | 57.18 |

It is to be noted that we have modelled the plant as a Single Input Single Output (SISO) system (between light stimulus and obtained electrical response) and ignored the impact of changing other factors such as temperature and humidity on plant electrophysiology, since their variation within a smaller time window is negligibly small. It is also well known that strongest correlation of the plant electrical response is with the rate of change of stimulus. Here the photosynthetic light intensity is widely varied as described in the experiment section which rapidly changes the polarization of the guard cells, thereby affecting $K^+$ ion concentration and as a result affecting the ionic current recorded. On contrary, a natural (un-manipulated) variation of temperature and humidity can always be considered to be constant during the experiment conducted over a short period of time [40]. There are also some theoretical arguments that at least a temperature change of 10°C is necessary to generate action potentials in plants as per the study by Sukhov and Vodeneev [23]. Since within half an hour duration of our experiment, the temperature change was negligible, it has not been considered in the modelling. In Table 5, the ambient temperature and humidity within the Faraday cage have also been reported for each experimental condition (which are taken from the original database), but due to their static nature they have not been considered as additional inputs for the modelling. Although there is a possibility of increase in leaf temperature if the light source is kept very close to the plant. Since, the light source used for the present experiments were LED which produces cool lighting, we have only measured the room temperature as an indicator of the ambience as well as the plant surface temperature. A larger deliberate variation in temperature (heat or cold shock) could be taken as an additional input to investigate thermal effects on such models in future research.

## 5. Conclusion

By using the electrical response data of 20 plants including three different species *viz. Laurus nobilis*, *Zamioculcas zamiifolia*, and *Cucumis sativus* to incident light stimulus, the rising and falling edges of the light has been successfully predicted within a forward and inverse modelling dynamical system framework. The best prediction for detecting the instants of turning on/off and peak intensity of light was obtained by a set of NLHW models over linear and NLARX models. The top three models show consistent prediction ability across all the datasets under different light pulse widths. It was also found that if the morphology of the electrical response is different from the response that is used to train the models, then the prediction accuracies deteriorate a bit. This method of system modelling, based on system identification approach, can be further explored by using plant electrical response data to determine a variety of stimulus such as introduction of gas, chemical to the soil etc., thus paving the path towards conceptualizing plant-based novel environmental biosensors in future.

**Acknowledgement:**

The work reported in this paper was supported by project PLants Employed As SEnsor Devices (PLEASED), EC grant agreement number 296582. The data on experiments on lights are available in the PLEASED website at http://pleased-fp7.eu/?page_id=253. We thank the anonymous reviewers for providing helpful comments and suggesting ways to increase the quality of the paper.